\documentclass[prl,twocolumn,amsmath,amssymb,floatfix,superscriptaddress]{revtex4-1}

\usepackage{graphicx}
\usepackage{amssymb}
\usepackage{amsmath}
\usepackage{color}
\usepackage{ wasysym }

\def\barray{\begin{array}}
\def\earray{\end{array}}
\def\be{\begin{equation}}
\def\ee{\end{equation}}
\def\ben{\begin{equation} \nonumber}
\def\een{\end{equation}}
\def\ban{\begin{eqnarray*}}
\def\ean{\end{eqnarray*}}
\def\ba{\begin{eqnarray}}
\def\ea{\end{eqnarray}}

\def\({\left(}
\def\){\right)}

\graphicspath{{./fig/}}

\begin{document}

\title{Warm (Pseudo)Scalar Inflation}

\author{ Vahid. Kamali}
\affiliation{Department of Physics, Bu-Ali Sina University, Hamedan 65178,
016016, Iran}
\affiliation{School of Physics,
Institute for Research in Fundamental Sciences (IPM),
19538-33511, Tehran, Iran}


\begin{abstract}
In this note, we will introduce an action for warm inflation model with direct coupling between the pseudoscalar field and mass-less  $SU(2)$ gauge fields. The potential of inflaton is protected against the thermal corrections in a thermal bath of gauge fields even with strong direct interaction between inflaton and light fields. The dissipation parameter of this model is approximately constant in the high-dissipative regime. In this regime, the model is compatible with observational data and non-Gaussianity is in order of Hubble slow-roll parameter  $\frac{\epsilon_{\phi}}{1+Q}$ even in $f<M_p$ limit.     
\end{abstract}
\maketitle

{\bf The set up and motivation:}
We can show that the inflation epoch occurs  at the finite temperature $T>H$. The model with interaction between inflaton and other (mainly) light fields is named warm inflation (WI) \cite{Berera:1995wh, Berera:1995ie}.
Out of warm inflation condition, i.e $T<H$, the nearly thermal bath of light particles cannot be sustained because of the accelerated expansion of the universe. In WI model there is an energy exchange between  the inflaton field and other light fields during slow-roll epoch, which leads to an additional friction term in the equation of motion (E.O.M) of inflaton scalar field at the background level.  
\begin{align}\label{E.O.M}
\ddot{\phi}+3H\dot{\phi}+\Upsilon\dot{\phi}+\frac{dV}{d\phi}=0
\end{align}
On the other hands when the light fields thermalize, the evolution of radiation energy density is presented by
\begin{align}\label{â€¢}
\dot{\rho}_R+4H\rho_R=\Upsilon\dot{\phi}^2
\end{align}
Transferring the energy density of inflaton field into the cosmic plasma may sustain the condition $T\geq H$ during inflation. In the slow-roll regime of radiation fluid:
\begin{align}\label{â€¢}
\dot{\rho}_R\ll 4H\rho_R \Rightarrow~~~~\rho_R=\frac{\Upsilon \dot{\phi}^2}{4H}=\frac{\pi^2}{30}g_{*}T^4
\end{align}
we can find 
\begin{align}\label{â€¢}
\frac{T}{H}=(\frac{\Upsilon}{H}\frac{\dot{\phi}^2}{H^4}\frac{15}{2g_{*}\pi^2})^{\frac{1}{4}}
\end{align}
If we have $\dot{\phi}\gg H^2,$ even for weak dissipative ($\Upsilon< H$) the main condition of warm inflation ($T>H$) can sustain in slow-roll regime ($\dot{\phi}\ll(V(\phi))^{\frac{1}{2}}\simeq H M_p$). Therefore the presence of dissipative effect may lead to WI rather than super-cold inflation. In the above equations, $g_{*}$ is the relativistic degree of freedom (DOF) and the source of radiation varies adiabatically when $\Upsilon=\Upsilon(\phi, T)$. The slow-roll conditions in WI model are modified 
\begin{align}\label{â€¢}
\epsilon_{\phi}& ~~\eta_{\phi}\ll 1+Q
\end{align}
where
\begin{align}\label{â€¢}
\epsilon_{\phi}=\frac{1}{2â€¢}M_p^2(\frac{V'}{V})^2~~~~\eta_{\phi}=M_p^2(\frac{V''}{V})~~~~~Q=\frac{\Upsilon}{3H}
\end{align} \cite{Berera:2008ar,BasteroGil:2009ec,Bastero-Gil:2016qru}. 
These new slow-roll conditions show the additional friction term $\Upsilon\dot{\phi}$ in modified E.O.M can alleviate the required flatness of the potential in the slow-roll regime. Using E.O.M \ref{E.O.M} of thermalized inflaton  we can find 
\begin{align}\label{â€¢}
\frac{\rho_R}{V(\phi)}\simeq\frac{1}{2â€¢}\frac{\epsilon_{\phi}}{1+Q}\frac{Q}{1+Q}
\end{align}
where 
\begin{align}\label{â€¢}
\ddot{\phi}<3H(1+Q)\dot{\phi}
\end{align}
which presents $\rho_R<V(\phi)$ during accelerated expansion epoch of the universe evolution (slow-roll inflation). But radiation energy density can smoothly become dominant component at the end of inflation, where $\epsilon_{\phi}\sim 1+Q$ with $Q\gg 1,$ without a need for a separate reheating epoch. Dissipation effects also modify the growth of inflaton fluctuations with distinctive imprint on the primordial spectrum. 

{\bf Problems of Warm Inflation:}
It was shown that the idea of WI not easy to present as a concrete model \cite{Yokoyama:1998ju}. Inflation field with direct coupling to light fields has problems. For example Yukawa interaction $\lambda\phi\bar{\psi}\psi$ leads to fermions mass $m_{\psi}=\lambda\phi$. Typically large inflaton values are needed in the slow-roll limit. But we want light fermions. On the other hands when small coupling $\lambda$ is considered the thermal bath condition $T>H$ may not be sustained. Direct coupling to light fields may add large thermal correction to inflaton mass $\sim \lambda T$, that can stop slow-roll condition $\dot{\phi}^2\ll V(\phi) $ in warm inflation regime $T>H$. In the literature, mostly WI has been explained by inflaton which indirectly couple to light DOF through heavy intermediate fields \cite{BasteroGil:2010pb,BasteroGil:2012cm}. Thermal correction of inflaton mass exponentially suppressed but there is a new problem in this realization: Dissipation coefficient is suppressed by powers $(\frac{T}{M_m})<1$ where $M_m$ is mediator mass.
 Solving this problem implies that we need a large number of intermediate fields which are required to hold a thermal bath for $50-60$ e-folds of inflation. The case of the brane constructions can be used for WI realization which is discussed in \cite{BasteroGil:2011mr}.
    
{\bf Warm Inflation with a few Light fields: }
Recently a new idea of WI has been published which named "warm little inflaton" (WLI) \cite{Bastero-Gil:2016qru}. In the WLI scenario, for the first time, WI may be realized by directly coupling Pseudo Nambu-Golden boson (PNGB) of broken gauge symmetry as inflaton field to a few light fields. The critical point in WLI is that Higgs boson is PNGB of a broken gauge symmetry breaking which has protected mass against large radiative corrections \cite{Schmaltz:2005ky}. In this note, we will introduce a new warm inflation model where inflaton directly coupled to light fields. In our scenario, Chern-Simons interaction between the inflaton field and non-Abelian gauge fields will be proposed: 

\begin{align}\label{action}
\mathcal{L}= \sqrt{-g}[-\frac{R}{2}-\frac{1}{4}F_{\mu\nu}^a F_a^{\mu\nu}-\frac{1}{2}\partial_{\mu}\partial^{\mu}\varphi-V(\varphi)-\frac{\varphi}{8M}F_{\mu\nu}^a\tilde{F}_a^{\mu\nu}]
\end{align}
where $M$ is symmetry breaking scale and $ \tilde{F}^{\mu\nu}=\frac{1}{2}\epsilon^{\mu\nu\rho\sigma}F_{a\mu\nu}$ is dual field strength \cite{Maleknejad:2011jw,Adshead:2013nka}  and 

\begin{align}\label{â€¢}
F^a_{\mu\nu}=\frac{1}{ig}[D_{\mu}^a,D_{\nu}^a]~~D^a_{\mu}=\partial_{\mu}-ig A_{\mu} J^{a}~~Tr[J_a,J_b]=\frac{1}{2}\delta_{ab}
\end{align}
The E.O.M for gauge field strength tensor $"F"$ in the presence of Chern-Simons like coupling is presented as:
   
\begin{align}\label{E.O.M F}
(\delta^{ab}\nabla_{\alpha}-g f^{abc}A_{\alpha}^c)F^{a\alpha\beta}-\frac{\epsilon^{\mu\nu\beta\alpha}}{2M}\partial_{\alpha}^{ab}(\phi F_{\mu\nu}^b)=0
\end{align}
where $\nabla_{\alpha}$ is the space-time covariant derivative of the inflaton field and $\phi=<\varphi>$ is the thermal average of the inflaton field $\varphi$. A thermal system with the finite temperature $T$ in our model is a cosmic expanding plasma which is composed of  inflaton with an equation of state (E.O.S) $w=\frac{P}{\rho}<0$ and gauge-light field particles with radiation like equation of state $w=\frac{P}{\rho}=\frac{1}{3}$
\footnote{There are two methods for covering isotropic symmetry for gauge fields
\cite{Golovnev:2008cf}:
 The first one is introducing a collective mode $A^a_{i}=\phi(t)\delta^a_i
 $ with a delta function between internal index and space index \cite{Adshead:2012kp,Maleknejad:2011sq}. The second one is 
considering a configuration of quantized random direction fields at various positions in the space. With averaging of these fields the space part of the gauge field is integrated out and also the isotropic symmetry recovered. In our model, we have used the second one.}.

Light particles are quanta of gauge field 
\begin{align}\label{ansatz}
A_{i}^a=\int \frac{d^3k}{(2\pi)^3}\varepsilon_{i}^a(k,t)e^{ik^{\mu}x_{\mu}}=\int \frac{d^3k}{(2\pi)^3}\delta_i^aJ_a\varepsilon(k,t)e^{ik^{\mu}x_{\mu}}~~~~~A_0=0
\end{align}
where $J_a$ is a generator of $SU(2)$ with commutation relations
\begin{align}\label{â€¢}
[J_a,J_b]=if_{abc}J_c
\end{align}
with $SU(2)$ structure function $f_{abc}=\epsilon_{abc}$.
Averaging on quantum fields leads to homogenous and isotropic E.O.M of gauge fields  (\ref{E.O.M F})
\begin{align}\label{E.O.M r}
\frac{\ddot{\Phi}}{a}+H\frac{\dot{\Phi}}{a}+2g^2\frac{\Phi^3}{a^3}=\frac{g}{2M}\dot{\phi}\frac{\Phi^2}{a^2}~~~~
\end{align}
where
\begin{align}\label{ansatz}
\Phi(t)\delta_i^a=a(t)\psi(t)\delta_i^a=<A_i^a>
\end{align}
The background evolution of warm pseudo-scalar (WPS) field in an isotropic and homogeneous thermal bath, 
 Using  (\ref{ansatz}) and action (\ref{action}), is presented by 
 \begin{align}\label{E.O.M in}
\ddot{\phi}+3H\dot{\phi}+\frac{dV}{d\phi}=-\frac{1}{2Ma^3}g\partial_t(\Phi^3)
\end{align}
In slow-roll limit of warm inflation $\frac{\ddot{\Phi}}{a}\ll H\frac{\dot{\Phi}}{a}$ (see the Appendix) and in $g\ll 1$ limit  the equations (\ref{E.O.M r})
\begin{align}\label{}
\frac{\dot{\Phi}}{a}\simeq\frac{g\psi^2}{2MH}\dot{\phi}~~~~
\end{align}
 and (\ref{E.O.M in})
\begin{align}\label{}
\ddot{\phi}+3H\dot{\phi}+\frac{dV}{d\phi}=-3H\frac{g\psi^2}{2MH}\frac{\dot{\Phi}}{a}
\end{align} 
  lead to E.O.M of warm inflation
\begin{align}\label{}
\ddot{\phi}+3H(1+Q)\dot{\phi}+\frac{d V}{d\phi}=0
\end{align}
where $Q=(\frac{g\psi^2}{2MH})^2>1,$ which is a function of background variables. In high dissipative regime the variation of  background variables  during last $60$ e-folds of inflation are very hard \cite{BasteroGil:2011xd}, so we have approximately constant dissipation parameter $Q$.   
In our model main properties  of warm inflation $(\frac{T}{H}>1~~~~\frac{\rho_r}{V(\phi)}<1)$ can be sustained. 
\begin{align}\label{â€¢}
\frac{\rho_r}{V}=\frac{\epsilon_{\phi}}{1+Q}\frac{Q}{1+Q}
\end{align}
where $Q>1$ \cite{SheikhJabbari:2012qf}. 

\begin{figure}
\begin{center}
\includegraphics[scale=0.25]{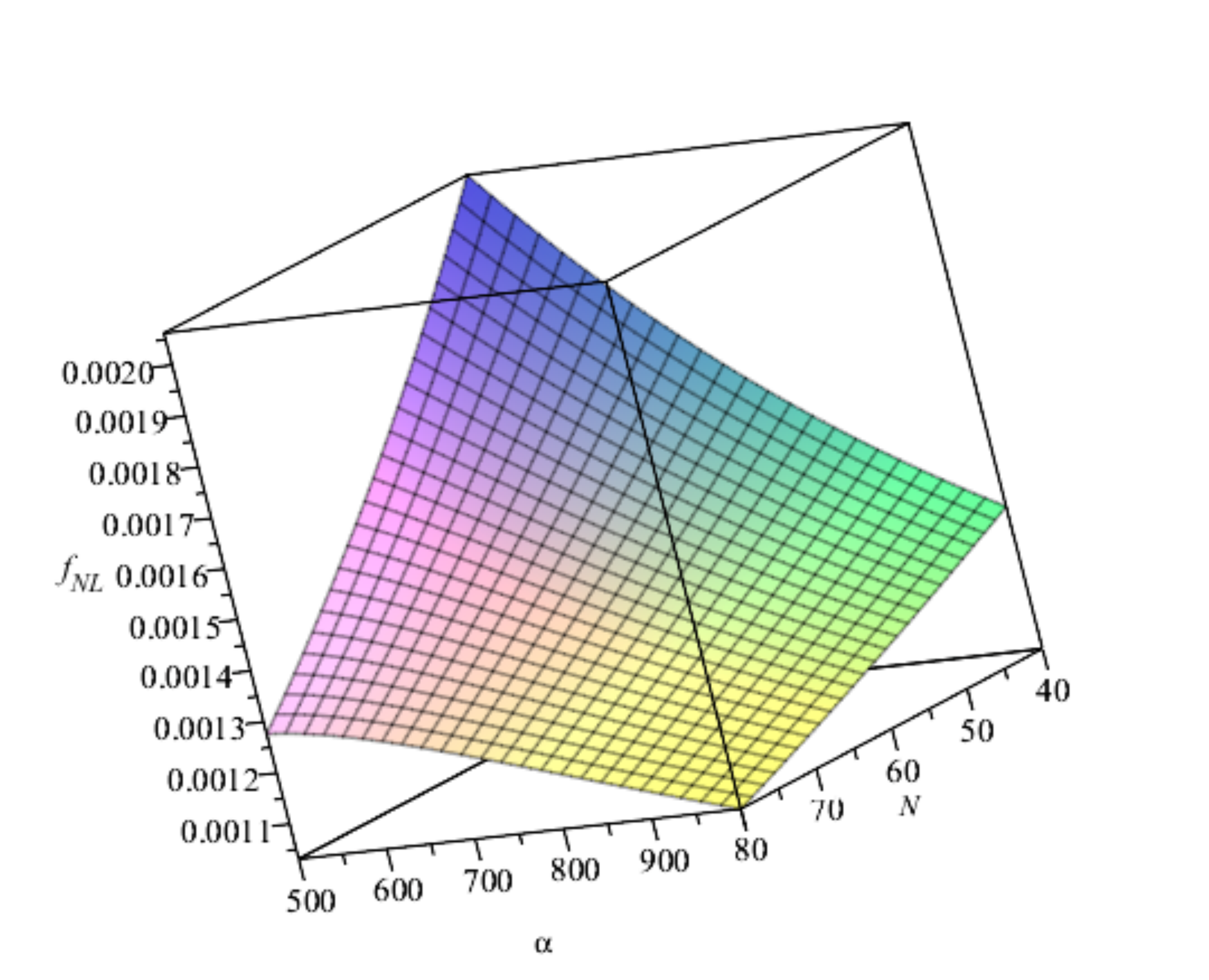}
\end{center}
\caption{Small values of non-Gaussianity are presented for reasonable values of $\alpha$}
\label{fNL}
\end{figure}

{\bf Perturbation:}
In warm inflation scenario the curvature power-spectrum is modified by dissipation effect:  
\begin{eqnarray}
\Delta^2_{\mathcal{R}}=\frac{V_{*}(1+Q_{*})^2}{24\pi^2M_p^4\epsilon_{\phi_{*}}}(1+2n_{*}+\frac{2\sqrt{3}\pi Q_{*}}{\sqrt{3+4\pi Q_{*}}}\frac{T_{*}}{H_{*}})G(Q)
\end{eqnarray}
The first term is power-spectrum of quantum fluctuations which is introduced in cold model of inflation. Another two terms are the modification of power spectrum in the warm scenario of inflation and $G$ is calculated numerically for temperature dependent dissipation parameters $Q$ \cite{Bastero-Gil:2016qru} . A well-known potential of pseudoscalar fields 
\begin{eqnarray}\label{Potential}
U(\phi)=m^2_{\phi}f^2(1+\cos(\frac{\phi}{f})),
\end{eqnarray}
is suitable for inflation epoch where $\frac{\phi}{f}$ is not close to $\pi$. We can study this potential in the context of warm pseudoscalar field model.
Leading terms of perturbation parameters in slow-roll  and high dissipative $Q>1$ limits are presented in table(\ref{tab:high})
\begin{table}[ht]
 \begin{center}
	\begin{tabular}{ | l | c | c |}
		\hline
		$Q> 1$ & Theoritical amount  & Constant parameter \\
		 \hline
		$\Delta_{\mathcal{R}}$ & $\Delta_{0\mathcal{R}}(\frac{U}{\epsilon_{\phi}})^{\frac{3}{2}}$ & $\Delta_{0\mathcal{R}}=\frac{Q^{\frac{9}{4}}}{2\sqrt{2}\pi(2\pi M_p^2)^{\frac{3}{2}}}$\\
		 \hline
		$n_s-1$ &  $-\frac{1.5}{N}-\frac{4.5}{\alpha}$& $\alpha=\frac{2(1+Q)f^2}{M_p^2},~~~\alpha>N$ \\
		\hline
		$r$ & $r_0\epsilon_{\phi}^{\frac{3}{4}}U^{\frac{1}{4}}$ & $r_0=\frac{16}{3\pi M_p^4 \Delta_{0\mathcal{R}}}$ \\
		\hline
		$\sin(\frac{\phi_{*}}{2f})$ &$A_0\exp(-\frac{3}{\alpha}N)$& $A_0=\cos(\frac{\sqrt{2}}{\alpha})$ \\
		\hline
	\end{tabular}
		\caption{Important perturbation parameters of pseudoscalar warm inflation in high dissipative regime $Q>1$  which are compared with observational data.}
		\label{tab:high}
 \end{center}
In figures (\ref{n-alpha}) and (\ref{r-n}) the best fitting of the model with observational data are presented for $Q>1$ and $f<M_p$ limits. 
\end{table}\\      
{\bf Non-Gaussianity:}
Now we consider the non-Gaussianity of our model using $\delta N$ formalism \cite{Zhang:2015zta}. In this method
perturbation theory of cosmology is studied by a quasi-homogeneous spatially-flat FLRW space-time with scale-factor $a(t)$. The spatial part of the FLRW metric is presented by:
\begin{eqnarray}
g_{ij}=a^2(t)e^{2\zeta(t,\textbf{x})}\gamma_{ij}(t,\textbf{x})
\end{eqnarray}  
where primordial curvature perturbation $\zeta$ is actually perturbation of $\ln(a)$ which has approximately Gaussian power-spectrum and $\gamma_{ij}$ is tensor perturbation. The nearly-Gaussian term $\zeta$ at least in a second order, which is good accuracy, is presented by derivatives of the number of e-folds w.r.t inflaton filed $\phi$:
\begin{eqnarray}
\zeta(t,\textbf{x})=N_{\phi}\delta\phi+\frac{1}{2}N_{\phi\phi}(\delta\phi)^2
\end{eqnarray} 
Perturbation of inflaton field $\delta\phi$ in warm inflation scenario is almost Gaussian which leads to scale independent 
non-linear parameter $f_{NL}$ as a function of derivatives of $N$ w.r.t inflaton filed \cite{Zhang:2015zta}:
\begin{eqnarray}
-\frac{3}{5}f_{NL}=\frac{1}{2}\frac{N_{\phi\phi}}{N_{\phi}^2}
\end{eqnarray}
where $N_{\phi}=-\frac{1}{M_p^2}\frac{U(1+Q)}{U_{\phi}}$. The non-linear parameter of non-Gaussianity parameter $f_{NL}$ in $Q\simeq const$ case is presented by
\begin{eqnarray}
f_{NL}=\frac{5}{6}\frac{1}{1+Q}(2\epsilon_{\phi}-\eta_{\phi})
\end{eqnarray}
which is in order of  Hubble slow-roll parameters. 
For our potential (\ref{Potential}) the level of non-Gaussianity is a function of the number of e-folds and parameter $\alpha$:
\begin{eqnarray}
f_{NL}=\frac{5}{3\alpha\cos^2(\frac{\sqrt{2}}{\alpha})}\exp(-\frac{6}{\alpha}N)
\end{eqnarray} 
Therefore, we have an insignificant non-Gaussianity of warm pseudo scalar inflation even with $f<M_p$. In figure (\ref{fNL}) the variations of non-gaussianity, during inflation, in term of parameter $\alpha$ is presented.

\begin{figure}
\begin{center}
\includegraphics[scale=0.25]{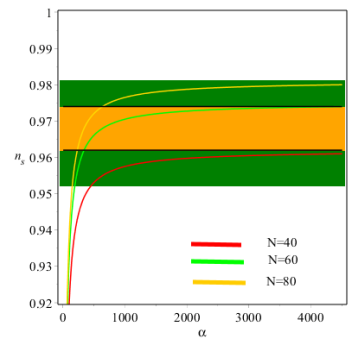}
\end{center}
\caption{Variation of $n_s$ in term of parameter $\alpha$ is presented for three cases of number of e-folds.}
\label{n-alpha}
\end{figure}

\begin{figure}
\begin{center}
\includegraphics[scale=0.25]{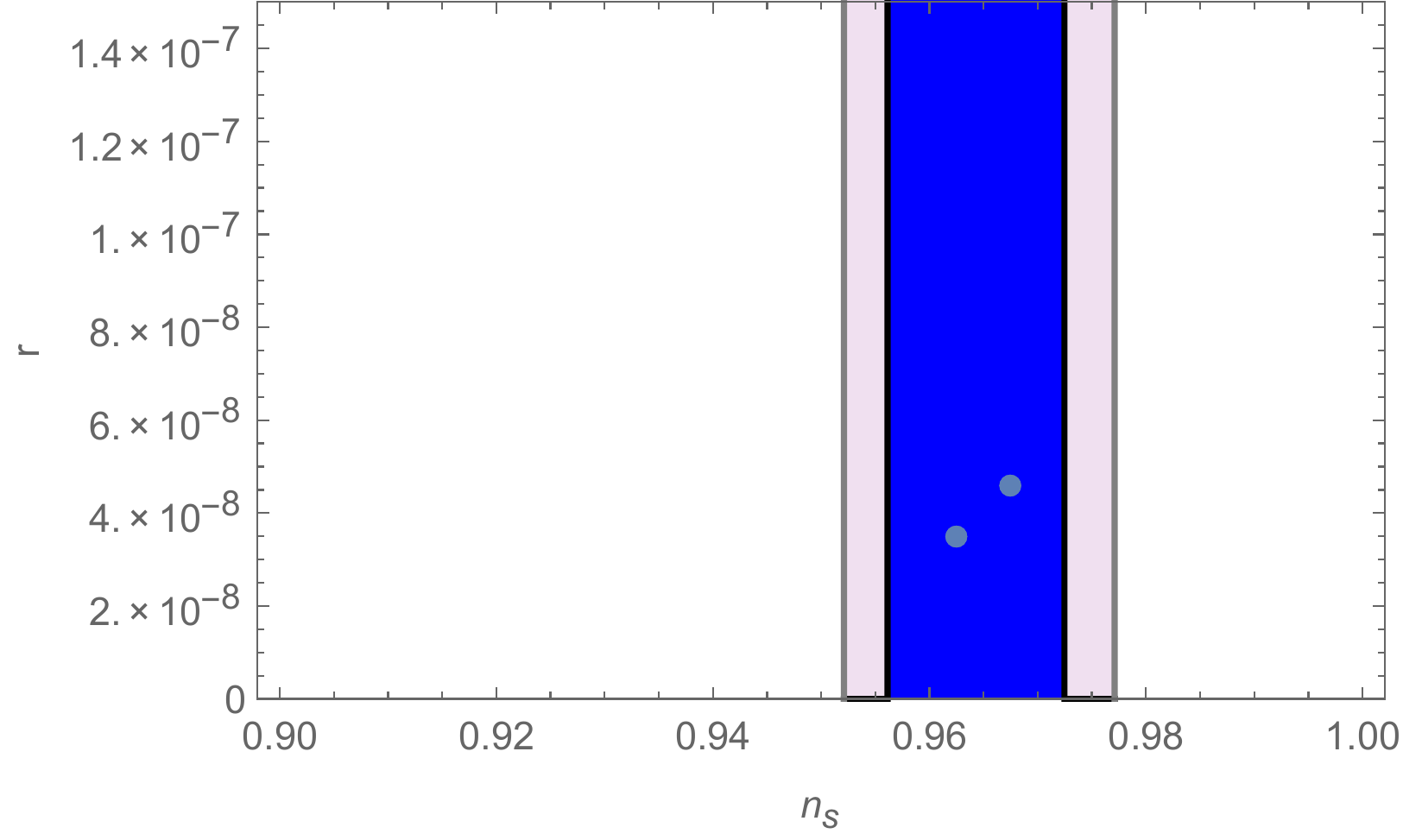}
\end{center}
\caption{$(r,n_s)$ of our model for two cases $N=60$ and $N=50$ is in $1-\sigma$ confidence level of $r-n_s$ planck results \cite{Akrami:2018odb}.}
\label{r-n}
\end{figure}

{\bf Observation constrain:}
In Fig.(\ref{n-alpha}), we show that the spectral index in term of parameter $\alpha$ for some amounts of the number of e-folds. The yellow region is $1\sigma$ confidence level of the spectral index. For $N=60$ case we have $450<\alpha<4500$ amounts which agree with observational data.    
Using $\alpha=600$ and two cases $N=50$ and $N=60$, we present two points in $1\sigma$ confidence level of $r-n_s$ graph (\ref{r-n}) where the upper point is the case $(\alpha=600,N=50)$ and next is $(\alpha=600,N=60)$ which are in good agreement with observational data with very small amount of tensor to scalar ratio.
In a phase space of model parameters $(\alpha,N)$ we show that our model has small amounts of non-Gaussianity (\ref{fNL}). 
 

{\bf Discussion and Conclusions:}
In this paper, we introduce a model of warm inflation with a direct connection between axion inflaton fields and light gauge fields. In the slow-roll limit, we have found approximately constant dissipation parameter $Q$ in term of background parameters of the model. The model is in good agreement with observational data in high dissipative regime $Q>1$ which can resolve the swampland conjecture \cite{Agrawal:2018own,Ooguri:2018wrx,Motaharfar:2018zyb,Bastero-Gil:2018yen}. The small amounts of non-Gaussianity are found for allowed phase space of the model parameters. 

{\bf Appendix }\label{slow}

In this section, we present the slow-roll condition of effective radiation field $\Phi$ and E.O.Ms by using energy conservation
\begin{eqnarray}\label{E.C}
\dot{\rho}+3H(\rho+P)=0\\
\nonumber
\rho=\rho_{\phi}+\rho_r\\
\nonumber
P=P_{\phi}+P_{r}
\end{eqnarray}
which leads to
\begin{eqnarray}
\dot{\rho}_{\phi}+3H(\rho_{\phi}+P_{\phi})=-[\dot{\rho}_r+3H(\rho_r+P_{r})
]
\end{eqnarray}
Using the action of our model we present 
\begin{eqnarray}
\rho_{\phi}=\frac{1}{2}\dot{\phi}^2+V(\phi)\\
\nonumber
P_{\phi}=\frac{1}{2}\dot{\phi}^2-V(\phi)
\end{eqnarray}
and 
\begin{eqnarray}
\rho_r=\frac{3}{2}(\frac{\dot{\Phi}^2}{a^2}+g^2\frac{\Phi^4}{a^4})\\
\nonumber
P_r=\frac{1}{2}(\frac{\dot{\Phi}^2}{a^2}+g^2\frac{\Phi^4}{a^4})
\end{eqnarray}
In warm inflation scenario, the slow-roll limit of radiation part is presented by $\dot{\rho}_r\ll 4H\rho_r$. Using the above equation we present 
\begin{eqnarray}
\dot{\rho}_{r}=\frac{3\dot{\Phi}}{a}(\frac{\ddot{\Phi}}{a}-H\frac{\dot{\Phi}}{a}+\frac{2g^2\Phi^3}{a^3})-\frac{6Hg^2\Phi^4}{a^4}
\end{eqnarray}

In slow-roll and $g\ll 1$ limits, we find the slow-roll condition for thermalizing  radiation field $\Phi$ as

\begin{eqnarray}
\frac{\ddot{\Phi}}{a}\ll 3H\frac{\dot{\Phi}}{a}
\end{eqnarray}
On the other hands, we can find LHS of the scalar field  and effective radiation field E.O.Ms using energy conservation (\ref{E.C})
\begin{eqnarray}
\dot{\phi}(\ddot{\phi}+3H\dot{\phi}+\frac{d V(\phi)}{d\phi})\\
\nonumber
=-\frac{3\dot{\Phi}}{a}(\frac{\ddot{\Phi}}{a}+H\frac{\dot{\Phi}}{a}+\frac{2 g^2\Phi^3}{a^3})
\end{eqnarray}
This relation agrees with E.O.Ms 
\begin{eqnarray}
\dot{\phi}[\ddot{\phi}+3H\dot{\phi}+\frac{dV}{d\phi}]=[-\frac{1}{a^3}\frac{g}{2M}\partial_t(\Phi^3)]\dot{\phi}\\
\nonumber
\frac{3\dot{\Phi}}{a}[\frac{\ddot{\Phi}}{a}+H\frac{\dot{\Phi}}{a}+2g^2\frac{\Phi^3}{a^3}]=[\frac{g}{2M}\dot{\phi}\frac{\Phi^2}{a^2}]\frac{3\dot{\Phi}}{a}
\end{eqnarray}
which come from the variation of effective action. We note that the RHS of E.O.Ms come from C.S interaction in the action.

\acknowledgements
I want to thank  Mohammad Mehdi Sheik-Jabbar,  Amjad Ashoorioon, Ali Akbar Abolhasani, 	Rudnei O. Ramos and J.G.Rosa. for some valuable discussions. 

 \bibliographystyle{apsrev4-1}
  \bibliography{ref}

\end{document}